\documentclass[english,11pt,a4paper]{article}

\usepackage[T1]{fontenc}
\usepackage[utf8]{inputenc}
\usepackage[english]{babel}

\usepackage{lmodern}
\usepackage[a4paper]{geometry}
\usepackage{bm}
\usepackage{amsmath}
\usepackage{amsthm}
\usepackage{amssymb}

\usepackage{mathtools}
\mathtoolsset{showonlyrefs}

\usepackage{cite}
\usepackage[title]{appendix}

\providecommand{\eprint}[1]{\href{https://arxiv.org/abs/#1}{arXiv:\nolinkurl{#1}}}

\usepackage{setspace}
\usepackage{authblk}

\usepackage{xcolor}

\usepackage[colorlinks=true,allcolors=blue,breaklinks=true,hypertexnames=false]{hyperref}

\author[$\dagger$]{Maciej Błaszak}
\author[$\ddagger$]{Krzysztof Marciniak\footnote{Corresponding author.}}
\author[$\dagger$]{Błażej M. Szablikowski}

\affil[$\dagger$]{Institute of Spintronics and Quantum Information\authorcr 
Faculty of Physics and Astronomy, Adam Mickiewicz University\authorcr
Uniwersytetu Pozna\'{n}skiego 2, 61-614 Poznań, Poland\authorcr
\texttt{blaszakm@amu.edu.pl, bszablik@amu.edu.pl}\authorcr
\texttt{\footnotesize ORCID: \href{https://orcid.org/0000-0002-3951-2850}{0000-0002-3951-2850},
\href{https://orcid.org/0000-0002-4001-6328}{0000-0002-4001-6328}}
\authorcr\mbox{}}

\affil[$\ddagger$]{Department of Science and Technology\authorcr
Link\"{o}ping University, Campus Norrk\"{o}ping\authorcr
601 74 Norrk\"{o}ping, Sweden\authorcr
\texttt{krzma@itn.liu.se}\authorcr
\texttt{\footnotesize ORCID: \href{https://orcid.org/0000-0003-3280-0160}{0000-0003-3280-0160}}}

\newtheorem{theorem}{Theorem}
\newtheorem{lemma}{Lemma}
\newtheorem{definition}{Definition}

\newtheorem{corollary}{Corollary}
\newtheorem{remark}{Remark}
\newtheorem{example}{Example}
\newtheorem{conclusion}{Conclusion}

\begin{document}
\title{\textsc{From pencils of Novikov algebras of Stäckel type to soliton hierarchies}}
\maketitle

\begin{abstract}
In this article we construct evolutionary soliton hierarchies from pencils of Novikov algebras of Stäckel type. We start by defining a special class of associative Novikov algebras, which we call Novikov algebras of Stäckel type, as they are associated with classical Stäckel metrics in Viète coordinates. We obtain sufficient conditions for pencils of these algebras so that the corresponding Dubrovin-Novikov Hamiltonian operators can be centrally extended, producing sets of pairwise compatible Poisson operators. These operators lead to coupled Korteweg-de~Vries (cKdV) and coupled Harry Dym (cHD) hierarchies, as well as to a triangular cKdV hierarchy and a triangular cHD hierarchy.
\end{abstract}

\bigskip

{\small\noindent\textbf{Keywords:} Novikov algebras, central extensions, soliton hierarchies,
multi-Hamiltonian structures 

}

\section{Introduction}

In this article we construct soliton hierarchies of evolutionary type from pencils of associative Novikov algebras of Stäckel type. Using this approach, we reconstruct the coupled Korteweg-de~Vries (cKdV) and coupled Harry Dym (cHD) hierarchies \cite{AF1,AF2,AF3} as well as the triangular cKdV and triangular cHD hierarchies.

There are various ways of constructing soliton hierarchies from appropriate algebraic structures. For example, in \cite{F1} the authors used loop algebras and $r$-matrix theory to produce compatible Poisson brackets leading to cKdV and cHD hierarchies. In \cite{B1} Frobenius algebras were applied to multi-component third-order local Poisson structures. In the article \cite{SS}, the authors performed the construction of $(1+1)$-dimensional integrable bi-Hamiltonian systems associated with Novikov algebras. The obtained systems were multi-component generalizations of the Camassa-Holm equation \cite{CH} that can be interpreted as Euler equations on the respective centrally extended Lie algebras. A similar approach for constructing multi-component soliton hierarchies, specifically Harry Dym and Hunter-Saxton, based on Frobenius triple, has been presented in \cite{Kon}.

The homogeneous first-order Hamiltonian operators \cite{GD,BN}, which are a special case of the Dubrovin-Novikov operators of hydrodynamic type \cite{DN}, have a very natural underlying algebraic structure. The conditions for a homogeneous operator
\begin{equation}\label{fho}
  \Pi^{ij} = \frac{1}{2}(b^{ij}_{k} + b^{ji}_{k}) u^k \frac{d}{dx} + \frac{1}{2}b^{ij}_{k} u^k_x,
\end{equation}
to be Hamiltonian are such that the $b^{ij}_{k}$ are the structure constants of a Novikov algebra \cite{BN}. Moreover, these operators can be defined through Lie-Poisson structures associated with the so-called translationally invariant Lie algebras, which are in one-to-one correspondence with Novikov algebras. For more information about this and directly related topics, see \cite{SS} and the recent works \cite{St1,St2,SZ}.

The associated translationally invariant Lie algebra can be centrally extended. The condition for the existence of cocycles (either first-order or third-order Gelfand-Fuks cocycles) is equivalent to the existence of symmetric bilinear forms on the Novikov algebra satisfying certain compatibility conditions (quasi-Frobenius and Frobenius). Second-order cocycles result in antisymmetric bilinear forms, which again satisfy certain algebraic relations \cite{BN,SS}.

In this article we introduce the concept of \emph{pencils of commutative Novikov algebras of Stäckel type} in order to construct centrally extended Poisson pencils of Dubrovin-Novikov type, which lead to soliton hierarchies of evolutionary type.

The article has the following structure. In Section \ref{S2} we review some known facts about Novikov algebras and central extensions of related Poisson operators. In Section \ref{S3} we consider particular associative Novikov algebras that we call of Stäckel type, as their first-order central extensions contain flat Stäckel metrics. In Section \ref{S4} we combine these single algebras into pencils of algebras of Stäckel type, which yield central extensions of Poisson pencils of Dubrovin-Novikov type, containing terms of first and third order. In the main theorem of this paper (Theorem \ref{Z1}) we establish sufficient conditions for the construction of such central extensions. These pencils in turn lead to soliton hierarchies of evolutionary type. In Section \ref{S5} we apply our theory to construct (i) the cKdV hierarchy, (ii) the cHD hierarchy (both in the convention used in \cite{AF2}), (iii) the triangular cKdV hierarchy, and (iv) the triangular cHD hierarchy.

\section{Novikov algebras, the associated Poisson operators and their central
extensions}\label{S2}

\begin{definition}
A finite-dimensional algebra $\mathbb{A}$ over $\mathbb{R}$ is called a
\emph{Novikov algebra} if it is right-commutative:
\begin{equation}
\left(  a\circ b\right)  \circ c    =\left(  a\circ c\right)  \circ b, \label{2.1a}
\end{equation}
and left-symmetric (quasi-associative):
\begin{equation}  
\left(  a\circ b\right)  \circ c-a\circ\left(  b\circ c\right)    =\left(
b\circ a\right)  \circ c-b\circ\left(  a\circ c\right). \label{2.1b}%
\end{equation}
Here $a,b,c\in\mathbb{A}$ and $\circ$ denotes the multiplication in $\mathbb{A}$.
\end{definition}

The quasi-associativity
condition implies that any non-commutative Novikov algebra $\mathbb{A}$ is Lie-admissible, that is, the
commutator $[a,b]=a\circ b-b\circ a$ defines the structure of a Lie algebra on
the underlying vector space $\mathbb{A}$.

Assume $\dim \mathbb{A}=n$ and let us choose a basis $e^{1},\ldots,e^{n}$ in $\mathbb{A}$. Let us denote the corresponding structure constants of the algebra $\mathbb{A}$  by $b_{k}^{ij}$. Thus\footnote{Throughout the paper we
use the Einstein summation convention, except in some cases where the summation symbol is used explicitly.}
\begin{equation}
  (a\circ b)_k  = b_k^{ij}a_ib_j\quad\text{or}\quad e^{i}\circ e^{j}=b_{k}^{ij}e^{k},   
\end{equation}
where $a,b\in\mathbb{A}$. Then, the corresponding $n\times n$ matrix $A$ with coefficients
\begin{equation}
  A^{ij}=b_{s}^{ij}e^{s}\label{mul}
\end{equation}
is the (multiplication) characteristic matrix of the algebra $\mathbb{A}$.

\begin{remark}
\label{jeden}If the algebra $\mathbb{A}$ is commutative
the structure constants of $\mathbb{A}$ are symmetric, that is $b_{k}^{ij}=b_{k}^{ji}$, while the conditions \eqref{2.1a} and \eqref{2.1b} reduce to the associativity condition
\[
\left(  a\circ b\right)  \circ c=a\circ\left(  b\circ c\right).%
\]
\end{remark}

For any Novikov algebra $\mathbb{A}$ we can consider the algebra
$\mathcal{L}_{\mathbb{A}}$ of all smooth $\mathbb{A}$-valued functions on
$x\in\mathbb{S}^{1}$. This algebra is equipped with the Lie bracket
\begin{equation}
\left[\!\left[  a,b\right]\!\right]  = a_{x}\circ b-b_{x}\circ a,\label{2.3}%
\end{equation}
where now $a,b\in\mathcal{L}_{\mathbb{A}}$ so that $a$ and $b$ depend on
$x\in\mathbb{S}^{1}$. Throughout the article we will use the letters $a,b,c,\ldots$
to denote elements of $\mathbb{A}$ as well as elements of $\mathcal{L}%
_{\mathbb{A}}$, which will be clear from the context. In
fact, the bracket \eqref{2.3} is a Lie bracket if and only if $\mathbb{A}$ is
a Novikov algebra \cite{BN}.

Consider now the following first-order operator
\begin{equation}
\Pi^{ij}
=\frac{1}{2}\left(b_{k}^{ij}+b_{k}^{ji}\right) q^{k}\frac{d}{dx}
+\frac{1}{2}b_{k}^{ij}q_{x}^{k},
\qquad i,j=1,\ldots,n,
\label{2.2}
\end{equation}
where $x\in\mathbb{S}^{1}$ and $q=(q^{1},\ldots,q^{n})$, with $q^{i}=q^{i}(x)$.
The operator \eqref{2.2} acts on $\mathcal{L}_{\mathbb{A}}$.
It is Poisson if and only if $b_{k}^{ij}$ are the structure constants of a
Novikov algebra \cite{GD,BN}.

The associated Poisson bracket is of Lie-Poisson type and is defined, for any
pair of functionals $\mathcal{H},\mathcal{F}$ on $\mathcal{L}_{\mathbb{A}}^{\ast}$, by
\begin{equation}
\{\mathcal{H},\mathcal{F}\}[q]
:=\int_{\mathbb{S}^{1}}
\frac{\delta\mathcal{H}}{\delta q^{i}}\,
\Pi^{ij}\,
\frac{\delta\mathcal{F}}{\delta q^{j}}\,dx
\equiv
\left\langle q,\left[\!\left[\delta_{q}\mathcal{H},\delta_{q}\mathcal{F}\right]\!\right]\right\rangle,
\qquad
q\in\mathcal{L}_{\mathbb{A}}^{\ast},\quad
\delta_{q}\mathcal{H},\delta_{q}\mathcal{F}\in\mathcal{L}_{\mathbb{A}},
\label{pb}
\end{equation}
with
\[
\delta_{q}\mathcal{H}=\frac{\delta\mathcal{H}}{\delta q^{i}}\,e^{i},
\qquad
\delta_{q}\mathcal{F}=\frac{\delta\mathcal{F}}{\delta q^{i}}\,e^{i}.
\]
The pairing between $\mathcal{L}_{\mathbb{A}}^{\ast}$ and $\mathcal{L}_{\mathbb{A}}$
is given by
\[
\langle q,a\rangle=\int_{\mathbb{S}^{1}}(q,a)\,dx,
\qquad
q\in\mathcal{L}_{\mathbb{A}}^{\ast},\quad a\in\mathcal{L}_{\mathbb{A}},
\]
where $(\cdot,\cdot)$ denotes the dual pairing between $\mathbb{A}^{\ast}$ and
$\mathbb{A}$.

Let us define a $2$-cocycle on $\mathcal{L}_{\mathbb{A}}$ as a bilinear form
$\omega:\mathcal{L}_{\mathbb{A}}\times\mathcal{L}_{\mathbb{A}}\rightarrow
\mathbb{R}$ such that $\omega$ is skew-symmetric:
\begin{equation}
\omega(a,b)=-\omega(b,a)\label{2.5a}%
\end{equation}
and satisfies the cyclic condition
\begin{equation}
\omega\left(\left[\!\left[a,b\right]\!\right],c\right)
+\omega\left(\left[\!\left[b,c\right]\!\right],a\right)
+\omega\left(\left[\!\left[c,a\right]\!\right],b\right)
=0.\label{2.5b}%
\end{equation}
With each such $2$-cocycle one can associate the following central extension of the Poisson bracket \eqref{pb}:
\begin{equation}
\left\{\mathcal{H},\mathcal{F}\right\}_{\omega}[q]
=\left\langle q,\left[\!\left[\delta_{q}\mathcal{H},\delta_{q}\mathcal{F}\right]\!\right]\right\rangle
+\omega\left(\delta_{q}\mathcal{H},\delta_{q}\mathcal{F}\right).
\label{2.6}%
\end{equation}
There are three types of differential $2$-cocycles on $\mathcal{L}_{\mathbb{A}}$
\cite{BN}.

A symmetric bilinear form $Z$ on $\mathbb{A}$ generates a $2$-cocycle \emph{of
order} $1$ on $\mathcal{L}_{\mathbb{A}}$ given by
\begin{equation}
\omega(a,b)=\int_{\mathbb{S}^{1}}Z\left(a_{x},b\right)\,dx\label{2.7}%
\end{equation}
if and only if the quasi-Frobenius condition
\begin{equation}
Z(a\circ b,c)=Z(a,c\circ b)\label{2.8}%
\end{equation}
is satisfied for any $a,b,c\in\mathbb{A}$. 
Such a cocycle yields the following extended Poisson operator:
\[
P^{ij}=\Pi^{ij}+Z^{ij}\frac{d}{dx},\qquad Z^{ij}=Z^{ji}.
\]

Further, an anti-symmetric bilinear form $Z$ on $\mathbb{A}$ generates a
$2$-cocycle \emph{of order} $2$ on $\mathcal{L}_{\mathbb{A}}$ given by
\[
\omega(a,b)=\int_{\mathbb{S}^{1}}Z\left(a_{xx},b\right)\,dx
\]
if and only if $Z$ satisfies the quasi-Frobenius condition \eqref{2.8} and,
additionally, the cyclic condition
\begin{equation}
Z(a\circ b,c)+Z(b\circ c,a)+Z(c\circ a,b)=0\label{cycl}%
\end{equation}
for all $a,b,c\in\mathbb{A}$. 

Notice that in the commutative case, for an anti-symmetric $Z$, the quasi-Frobenius condition \eqref{2.8} together with the cyclic condition  \eqref{cycl} reduce to the single
condition of the form
\begin{equation}\label{war}
  Z(a\circ b, c)=0,
\end{equation}
where $a,b,c\in\mathbb{A}$ are arbitrary. This is due to the fact that in this situation \eqref{cycl} reads
\begin{equation}
0=Z(a\circ b,c) - Z(a\circ b,c)-Z(a\circ b,c)=-Z(a\circ b,c)
\end{equation}
so \eqref{war} follows. This cocycle yields the following extended Poisson
operator:
\[
P^{ij}=\Pi^{ij}+Z^{ij}\frac{d^{2}}{dx^{2}},\qquad Z^{ij}=-Z^{ji}.
\]

Finally, a symmetric bilinear form $Z$ on $\mathbb{A}$ generates a $2$-cocycle
of order $3$ on $\mathcal{L}_{\mathbb{A}}$ given by
\begin{equation}
\omega(a,b)=\int_{\mathbb{S}^{1}}Z\left(a_{xxx},b\right)\,dx\label{2.10}%
\end{equation}
if and only if $Z$ satisfies the quasi-Frobenius condition \eqref{2.8} and,
additionally, the condition
\begin{equation}
Z(a,b\circ c)=Z(a,c\circ b).\label{2.11}%
\end{equation}
This cocycle yields the following extended Poisson operator:
\[
P^{ij}=\Pi^{ij}+Z^{ij}\frac{d^{3}}{dx^{3}},\qquad Z^{ij}=Z^{ji}.
\]

Note that in the commutative case the condition \eqref{2.11} is always
satisfied. Therefore, in this case the conditions for cocycles of order $1$
and order $3$ coincide and are given by the same quasi-Frobenius condition
\eqref{2.8}, which can be written as the (standard) Frobenius condition
\begin{equation}
Z(a\circ b,c)=Z(a,b\circ c),\qquad a,b,c\in\mathbb{A},\label{frob}
\end{equation}
or, equivalently, as
\begin{equation}\label{2.9}
Z(e^{i}\circ e^{j},e^{k})=Z(e^{i},e^{j}\circ e^{k}),\qquad i,j,k=1,\ldots,n. 
\end{equation}
Let $Z^{ij}:=Z(e^{i},e^{j})$. Then the Frobenius condition \eqref{frob}
reduces in coordinates to the following homogeneous system of linear equations
for the symmetric form $Z^{ij}=Z^{ji}$:
\begin{equation}
b_{s}^{ij} Z^{sk} - Z^{is} b_{s}^{jk} = 0,\qquad i,j,k=1,\ldots,n.
\label{3.3}
\end{equation}

Moreover, since the conditions \eqref{2.5a} and \eqref{2.5b} are linear in
$\omega$, an arbitrary linear combination of the above cocycles leads to a
corresponding centrally extended Poisson operator $P^{ij}$ as well.

\section{Novikov algebras of Stäckel type}\label{S3}

Consider a family of $n$-dimensional algebras $\mathcal{A}^m = (\mathbb{R}^m,\circ_m)$, defined for each
$m\in\{0,\ldots,n\}$, with the multiplication
\begin{equation}\label{algm}
  e^{i}\circ_m e^{j} =
  \begin{cases}
    e^{\,i+j+m-n-1}, & \text{for } i,j\in\{1,\ldots,n-m\}\equiv I_1^{m},\\[2pt]
    -\,e^{\,i+j+m-n-1}, & \text{for } i,j\in\{n-m+1,\ldots,n\}\equiv I_2^{m},\\[2pt]
    0, & \text{otherwise}.
  \end{cases}
\end{equation}
Thus the structure constants of $\mathcal{A}^m$ are given by
\begin{equation}
e^{i}\circ_m e^{j}=\left(b_{m}\right)_{s}^{ij}\,e^{s},\qquad
\left(b_{m}\right)_{s}^{ij}=
\begin{cases}
\delta_{s}^{\,i+j+m-n-1}, & i,j\in I_{1}^{m},\\
-\delta_{s}^{\,i+j+m-n-1}, & i,j\in I_{2}^{m},\\
0, & \text{otherwise}.
\end{cases}
\label{3.1}%
\end{equation}
As $\left(b_{m}\right)_{s}^{ij}=\left(b_{m}\right)_{s}^{ji}$ in \eqref{3.1},
it follows that every $\mathcal{A}^{m}$ is commutative. In fact, each
$\mathcal{A}^{m}$ is a Novikov algebra, since the following assertion holds
(cf.\ Remark~\ref{jeden}).

\begin{lemma}
\label{L1}
All $\mathcal{A}^{m}$ are associative.
\end{lemma}

This lemma is a straightforward consequence of Remark~\ref{rem} below.
The algebra $\mathcal{A}^{m}$ will henceforth be
called the $m$-th Novikov algebra of Stäckel type, due to considerations below. Moreover, $\mathcal{A}^{n}$ is the only one among the $\mathcal{A}^{m}$ that has a
unity element, namely $-e^{1}$.

\begin{example}
\label{P1}
For $n=4$, the multiplication matrices $A_m$ defined in \eqref{mul} by
$(A_m)^{ij}=(b_m)^{ij}_{s}\,e^{s}$, with the structure constants \eqref{3.1}, are
\[
A_{0} =
\begin{pmatrix}
0 & 0 & 0 & 0\\
0 & 0 & 0 & e^{1}\\
0 & 0 & e^{1} & e^{2}\\
0 & e^{1} & e^{2} & e^{3}
\end{pmatrix},
\quad
A_{1}=
\begin{pmatrix}
0 & 0 & 0 & 0\\
0 & 0 & e^{1} & 0\\
0 & e^{1} & e^{2} & 0\\
0 & 0 & 0 & -e^{4}
\end{pmatrix},
\quad
A_{2}=
\begin{pmatrix}
0 & 0 & 0 & 0\\
0 & e^{1} & 0 & 0\\
0 & 0 & -e^{3} & -e^{4}\\
0 & 0 & -e^{4} & 0
\end{pmatrix},
\]
\[
A_{3} =
\begin{pmatrix}
0 & 0 & 0 & 0\\
0 & -e^{2} & -e^{3} & -e^{4}\\
0 & -e^{3} & -e^{4} & 0\\
0 & -e^{4} & 0 & 0
\end{pmatrix},
\quad
A_{4}=
\begin{pmatrix}
-e^{1} & -e^{2} & -e^{3} & -e^{4}\\
-e^{2} & -e^{3} & -e^{4} & 0\\
-e^{3} & -e^{4} & 0 & 0\\
-e^{4} & 0 & 0 & 0
\end{pmatrix}.
\]
\end{example}

\begin{remark}\label{rem}
Note that the algebra $\mathcal{A}^n$ can be represented as $n$-dimensional algebra of truncated polynomials
\begin{equation}\label{pol}
  \mathbb{I}_n \cong \mathbb{R}[x]/\langle x^{n} \rangle,\qquad e^{i}\circ e^{j} = -e^{i+j-1},
\end{equation}
where $e^{i}= -x^{i-1}$, for $i=1,\ldots,n$, are basis vectors. Similarly, the algebra $\mathcal{A}^0$ can be represented as
the subalgebra of truncated polynomials without free (constant) terms
\begin{equation}\label{polt}
  \tilde{\mathbb{I}}_n \cong x\mathbb{R}[x]/\langle x^{n+1} \rangle,\qquad e^{i}\circ e^{j} = e^{i+j-n-1},
\end{equation}
where $e^{i}= x^{n+1-i}$, for $i=1,\ldots,n$, are basis vectors. The above algebras are obviously commutative and associative. Note that $\tilde{\mathbb{I}}_1$ is a trivial algebra. Then, all the $n$-dimensional algebras from the family $\mathcal{A}^m$ have the following structure:
\begin{equation}
  \mathcal{A}^0 \equiv \tilde{\mathbb{I}}_n,\qquad
  \mathcal{A}^m \cong \tilde{\mathbb{I}}_{n-m}\oplus \mathbb{I}_m\quad\text{for}\quad m\in\{1,\ldots,n-1\},\qquad
  \mathcal{A}^n \equiv \mathbb{I}_n.
\end{equation}
The minus sign in the definitions \eqref{algm} and \eqref{pol} is not merely a matter of convention,
but it is required by compatibility conditions, see Section~\ref{S4}.
\end{remark}
 In the case of our Novikov algebras of Stäckel type it is possible to find a general solution $Z$ of the Frobenius condition \eqref{3.3}.

\begin{theorem}\label{Z}
The general $n$-parameter solution of the Frobenius condition \eqref{3.3}
for the symmetric bilinear form $Z_m$ on the $m$-th algebra $\mathcal{A}^m$,
defined by \eqref{algm}, is given by
\begin{equation}
(Z_{m})^{ij}=
\begin{cases}
\ \ \varphi_{m}^{i+j+m-n-1},
\quad i,j\in I_{1}^{m},\\
-\varphi_{m}^{i+j+m-n-1},
\quad i,j\in I_{2}^{m},\\
\ \ 0, \quad \text{otherwise},
\end{cases}
\label{3.4}%
\end{equation}
where $(Z_m)^{ij}=Z_m(e^{i},e^{j})$ and $\varphi_{m}^{s}$ are
arbitrary real constants. 
\end{theorem}
Here and in what follows we use the notation $\varphi_{m}^{i}=0$ for $i<0$ and for $i>n$. The proof is given in the Appendix. 
Thus, each form $Z_{m}\equiv Z_{m}(\varphi)$
depends on $n$ parameters
$\varphi_{m}^{0},\ldots,\varphi_{m}^{n-m-1}$, $\varphi_{m}^{n-m+1},\ldots,\varphi_{m}^{n}$
and is explicitly given by
\begin{equation}
Z_{m}(\varphi)=\left(\!%
\begin{array}{c|c}
\begin{matrix}
 &  & \varphi_{m}^{0}\\
 & \reflectbox{$\ddots$} & \vdots\\
\varphi_{m}^{0} & \cdots & \varphi_{m}^{n-m-1}
\end{matrix}
& 0_{(n-m)\times m}\\ \hline
0_{m\times(n-m)} &
\begin{matrix}
-\varphi_{m}^{n-m+1} & \cdots & -\varphi_{m}^{n}\\
\vdots & \reflectbox{$\ddots$} & \\
-\varphi_{m}^{n} &  & 
\end{matrix}
\end{array}
\!\right),\qquad m=0,\ldots,n.
\label{Zmfi}%
\end{equation}

\begin{lemma}\label{L7}
\bigskip No $\mathcal{A}^{m}$ has a $2$-cocycle of order $2$.
\end{lemma}
The proof of this lemma is in the Appendix. 

Consequently, the Poisson operator corresponding to each algebra 
$\mathcal{A}^m$
\begin{equation}
\Pi_{m}^{ij}=(b_{m})_{k}^{ij}q^{k}\frac{d}{dx}+\frac{1}{2}(b_{m})_{k}%
^{ij}q_{x}^{k}\label{3.45}%
\end{equation}
(cf.\ \eqref{2.2}) can be centrally extended to the $2n$-parameter Poisson
operator%
\begin{equation}
P_{m}^{ij}= \Pi_{m}^{ij} +(Z_{m}%
)^{ij}(\varphi)\frac{d}{dx} + (Z_{m})^{ij}(\psi)\frac{d^{3}}{dx^{3}},\label{3.47}%
\end{equation}
which in the matrix form can be presented as
\begin{equation}
P_{m}=G_{m}(q,\varphi)\frac{d}{dx}+\frac{1}%
{2}\left[  G_{m}(q,\varphi)\right]  _{x} + Z_{m}(\psi)\frac{d^{3}}{dx^{3}},\label{3.5}%
\end{equation}
where $Z_{m}(\psi)$ is defined by \eqref{Zmfi} but with a new set of $n$ parameters
$\psi_{m}^{i}$, while
\begin{equation}
 G_{m}^{ij}(q,\varphi) := (b_m)^{ij}_k q^k + (Z_m)^{ij}(\varphi),
\end{equation}
so that
{\small\[
G_{m}(q,\varphi) =\left(\!%
\begin{array}{c|c}
\begin{matrix}
 &  &  &\varphi_{m}^{0}\\
 & & \reflectbox{$\ddots$} & q^1 + \varphi_{m}^{1}\\
 & \reflectbox{$\ddots$} & \reflectbox{$\ddots$} & \vdots\\
\varphi_{m}^{0} & q^1 + \varphi_{m}^{1} & \cdots & q^{n-m-1} + \varphi_{m}^{n-m-1}
\end{matrix}
& 0_{(n-m)\times m}\\ \hline
0_{m\times(n-m)} &
\begin{matrix}
-q^{n-m+1}-\varphi_{m}^{n-m+1} & \cdots & -q^n-\varphi_{m}^{n}\\
\vdots & \reflectbox{$\ddots$} & \\
-q^n-\varphi_{m}^{n} &  & 
\end{matrix}
\end{array}
\!\right).
\]}
From now on, $G_{m}\equiv G_{m}(q,\varphi)$ will be considered as a flat
contravariant metric on a pseudo-Euclidean space with coordinates
$(q^{1},\ldots,q^{n})$.

Note that \eqref{3.5} is the most general differential central extension of the Poisson
bracket \eqref{3.45}. Also, for a fixed $m$, the shift
\begin{equation}
q^{i}+\varphi_{m}^{i}\mapsto q^{i},\qquad i=1,\ldots,n,
\label{shifts}%
\end{equation}
transforms $G_{m}$ to the form
{\small\[
G_{m}(q,\varphi) =\left(\!%
\begin{array}{c|c}
\begin{matrix}
 &  &  & \varphi_{m}\\
 &  & \reflectbox{$\ddots$} & q^{1}\\
 & \reflectbox{$\ddots$} & \reflectbox{$\ddots$} & \vdots\\
\varphi_{m} & q^{1} & \cdots & q^{n-m-1}
\end{matrix}
& 0_{(n-m)\times m}\\ \hline
0_{m\times(n-m)} &
\begin{matrix}
-q^{n-m+1} & \cdots & -q^{n}\\
\vdots & \reflectbox{$\ddots$} & \\
-q^{n} &  &
\end{matrix}
\end{array}
\!\right).\label{3.6}
\]}
Here we denote the only remaining parameter $\varphi_{m}^{0}$ by $\varphi_{m}$.
This means that the first-order central extension of the Poisson operator
\eqref{3.45} is $1$-parameter for $m=0,\ldots,n-1$ and trivial for $m=n$.

\begin{example}
\label{P2}
For $n=4$ the matrices $G_{m}$ in \eqref{3.6} take the form
\[
G_{0} =
\begin{pmatrix}
0 & 0 & 0 & \varphi_{0}\\
0 & 0 & \varphi_{0} & q^{1}\\
0 & \varphi_{0} & q^{1} & q^{2}\\
\varphi_{0} & q^{1} & q^{2} & q^{3}
\end{pmatrix},
\quad
G_{1}=
\begin{pmatrix}
0 & 0 & \varphi_{1} & 0\\
0 & \varphi_{1} & q^{1} & 0\\
\varphi_{1} & q^{1} & q^{2} & 0\\
0 & 0 & 0 & -q^{4}
\end{pmatrix},\quad
G_{2} =
\begin{pmatrix}
0 & \varphi_{2} & 0 & 0\\
\varphi_{2} & q^{1} & 0 & 0\\
0 & 0 & -q^{3} & -q^{4}\\
0 & 0 & -q^{4} & 0
\end{pmatrix},
\]
\[
G_{3} =
\begin{pmatrix}
\varphi_{3} & 0 & 0 & 0\\
0 & -q^{2} & -q^{3} & -q^{4}\\
0 & -q^{3} & -q^{4} & 0\\
0 & -q^{4} & 0 & 0
\end{pmatrix},\quad
G_{4} =
\begin{pmatrix}
-q^{1} & -q^{2} & -q^{3} & -q^{4}\\
-q^{2} & -q^{3} & -q^{4} & 0\\
-q^{3} & -q^{4} & 0 & 0\\
-q^{4} & 0 & 0 & 0
\end{pmatrix}.
\]
\end{example}

Each of the metrics $G_{m}$ in \eqref{3.6} attains, after the rescaling $q_{i}' =\varphi_{m}^{-1}q_{i}$, and assuming that $\varphi_{m}\neq 0$ (this rescaling is not necessary for $G_n$), the form of a Stäckel metric in Viète coordinates \cite{BS}. Further transformation from Viète coordinates to
separation coordinates $(\lambda_{1},\ldots,\lambda_{n})$ takes the form
\[
q_{i}'=(-1)^{i} s_{i}(\lambda_{1},\ldots,\lambda_{n}),\qquad i=1,\ldots,n,
\]
where $s_{i}(\lambda_{1},\ldots,\lambda_{n})$ are the elementary symmetric polynomials.
In the separation coordinates $(\lambda_{1},\ldots,\lambda_{n})$ the metric $G_m$ attains the diagonal form
\begin{equation}
G_{m}=\varphi_{m}^{-1}
\begin{pmatrix}
\frac{\lambda_{1}^{m}}{\Delta_{1}} &  & 0\\
 & \ddots & \\
0 &  & \frac{\lambda_{n}^{m}}{\Delta_{n}}
\end{pmatrix},
\qquad
\Delta_{i}=\prod_{j\neq i}(\lambda_{i}-\lambda_{j}).
\label{3.7}%
\end{equation}

\section{Novikov pencils of Stäckel type}\label{S4}

Let us now define the algebra $\mathcal{A}$ with the multiplication $\circ$
given by arbitrary linear combinations of the multiplications from the algebras
$\mathcal{A}^m$:
\begin{equation}\label{4.1}%
e^{i}\circ e^{j} := \sum_{m=0}^n \alpha_m\, e^{i}\circ_m e^{j},
\end{equation}
where the parameters $\alpha_m\in \mathbb{R}$. The associated structure constants
are
\begin{equation}\label{4.0}
b_{s}^{ij}=\sum_{m=0}^{n}\alpha_{m}(b_{m})_{s}^{ij},
\qquad
e^{i}\circ e^{j} = b^{ij}_{s} e^{s}.
\end{equation}
The algebra $\mathcal{A}$ is commutative and below we show that it is still
associative. Hence it is also a Novikov algebra, depending now on
$n+1$ arbitrary parameters $\alpha^{m}$, $m=0,\ldots,n$. We will call the algebra $\mathcal{A}$ a Novikov pencil of Stäckel type.

\begin{lemma}
\label{L2}
The algebra $\mathcal{A}$ is associative for arbitrary choice of the parameters $\alpha_m$.

Equivalently, the multiplications from the associative algebras $\mathcal{A}^m$
are mutually compatible, i.e.
\begin{equation}\label{compat}
(a\circ_m b)\circ_p c + (a\circ_p b)\circ_m c
= a\circ_m (b\circ_p c) + a\circ_p (b\circ_m c),
\end{equation}
for all $a,b,c\in \mathcal{A}$ and $m,p=0,\ldots,n$.
\end{lemma}

The proof is in the Appendix.

\begin{conclusion}
The operator%
\[
\Pi=\sum_{m=0}^{n}\alpha_{m}\Pi_{m},
\]
where $\Pi_{m}$ are $n+1$ Poisson operators of the form \eqref{3.45},
is Poisson for all values of $\alpha_{m}$, so that all $\Pi_{m}$ are pairwise compatible.
\end{conclusion}

Due to the commutativity of $\mathcal{A}$, the Frobenius condition for the Novikov pencil $\mathcal{A}$ has also the form  \eqref{2.9} with the multiplication $\circ$ defined by \eqref{4.1}. The theorem below shows that in this situation there exists a particular solution $Z$ of
\eqref{2.9} that has the form of a pencil of all $Z_{m}$ with exactly the same
coefficients $\alpha_{m}$ as in the Novikov pencil \eqref{4.0}.

\begin{theorem}
\label{Z1}
The pencil%
\begin{equation}
Z=\sum_{m=0}^{n}\alpha_{m}Z_{m},\label{4.4}%
\end{equation}
where the bilinear forms $Z_{m}$
are given by \eqref{3.4}, satisfies the Frobenius condition \eqref{frob}
on the algebra $\mathcal{A}$ defined by \eqref{4.1}, for any choice of the parameters $\alpha_m$, provided that%
\begin{equation}
\varphi_{m}^{s}=\varphi^{s},\qquad s=0,\ldots,n.\label{4.5}%
\end{equation}
\end{theorem}
The condition \eqref{4.5} means that all the bilinear forms $Z_{m}$ in
\eqref{4.4} share the same set of parameters. Explicitly, the bilinear forms $Z_{m}$ in the pencil \eqref{4.4} have the form
\begin{equation}
(Z_{m})^{ij}=
\begin{cases}
\ \ \varphi^{i+j+m-n-1},
\quad i,j\in I_{1}^{m},\\
-\varphi^{i+j+m-n-1},
\quad i,j\in I_{2}^{m},\\
\ \ 0, \quad \text{otherwise},
\end{cases}
\label{4.6}%
\end{equation}
so that $Z$ depends on $n+1$ parameters $\varphi^{0},\ldots,\varphi^{n}$,
and each $Z_{m}$ depends on the same parameters except for $\varphi^{n-m}$.

The proof is in the Appendix.

\begin{remark}\label{Bl}
The Frobenius condition \eqref{2.9} is equivalent to demanding
\begin{equation}
  Z_m(e^i\circ_p e^j,e^k) + Z_p(e^i\circ_m e^j,e^k)
  = Z_m(e^i, e^j\circ_p e^k) + Z_p(e^i, e^j\circ_m e^k),
  \qquad i,j,k=1,\ldots,n,
\end{equation}
for all pairs $m,p=0,\ldots,n$.
\end{remark}

\begin{corollary}
The operator $P=\sum\limits_{m=0}^{n}\alpha_{m}P_{m}$ is Poisson, with
\begin{equation}
P_{m}=Z_{m}(\psi^{0},\ldots,\psi^{n})\frac{d^{3}}{dx^{3}}+\Pi_{m}
+Z_{m}(\varphi_{0},\ldots,\varphi_{n})\frac{d}{dx},\qquad m=0,\ldots,n.
\label{4.6a}%
\end{equation}
Here $\psi^{0},\ldots,\psi^{n}$ and $\varphi_{0},\ldots,\varphi_{n}$ are two
independent sets of parameters used to construct the first- and third-order
central extensions by \eqref{4.6}, and thus all $P_{m}$ are pairwise compatible.
\end{corollary}

Moreover, since $\varphi_{m}^{s}$ do not depend on $m$ (cf. \eqref{4.5}),
there exists a common shift (cf. shifts \eqref{shifts})
\[
q^{i}+\varphi^{i}\rightarrow q^{i},\qquad i=1,\ldots,n,
\]
that turns all $P_{m}$ in \eqref{4.6a} into the following $(n+1)$-parameter
form:
\begin{equation}
P_{m}=Z_{m}(\psi^{0},\ldots,\psi^{n})\frac{d^{3}}{dx^{3}}+\Pi_{m}
+Z_{m}(\varphi)\frac{d}{dx},\qquad m=0,\ldots,n,
\label{4.8}%
\end{equation}
where
\[
Z_{m}(\varphi)=\left(\!\renewcommand\arraystretch{1.3}%
\begin{array}{c|c}%
\begin{matrix}
 &  & \varphi\\
 & \reflectbox{$\ddots$} & \\
\varphi &  & 
\end{matrix}
& 0_{(n-m)\times m}\\\hline
0_{m\times(n-m)} & 0_{m\times m}%
\end{array}
\!\right),\qquad m=0,\ldots,n,
\]
(with $\varphi_{0}=\varphi$, note that $Z_{n}=0$) and
{\small\[
Z_{m}(\psi_0,\ldots,\psi_n)=\left(\!\renewcommand\arraystretch{1.3}%
\begin{array}{c|c}%
\begin{matrix}
 &  & \psi^{0}\\
 & \reflectbox{$\ddots$} & \vdots\\
\psi^{0} & \cdots & \psi^{n-m-1}
\end{matrix}
& 0_{(n-m)\times m}\\\hline
0_{m\times(n-m)} &
\begin{matrix}
-\psi^{n-m+1} & \cdots & -\psi^{n}\\
\vdots & \reflectbox{$\ddots$} & \\
-\psi^{n} &  & 
\end{matrix}
\end{array}
\!\right),\qquad m=0,\ldots,n.
\]}
Thus, $(n+1)$ compatible Poisson operators of third order \eqref{4.8} can be
written in a compact form
{\small 
\begin{equation}
P_{m}=\left(\!\renewcommand\arraystretch{1.3}%
\begin{array}{c|c}%
\begin{matrix}
 &  & j_{0}\\
 & \reflectbox{$\ddots$} & \vdots\\
j_{0} & \cdots & j_{n-m-1}
\end{matrix}
& 0_{(n-m)\times m}\\\hline
0_{m\times(n-m)} &
\begin{matrix}
-j_{n-m+1} & \cdots & -j_{n}\\
\vdots & \reflectbox{$\ddots$} & \\
-j_{n} &  & 
\end{matrix}
\end{array}
\!\right),\qquad m=0,\ldots,n,
\label{4.9}
\end{equation}
}
where
\[
j_{0}=\varphi \frac{d}{dx}+\psi^{0}\frac{d^{3}}{dx^{3}},\qquad
j_{k}=\frac{1}{2}\left(q_{k} \frac{d}{dx}+ \frac{d}{dx} q_{k}\right)+\psi^{k} \frac{d^{3}}{dx^{3}},
\qquad k=1,\ldots,n.
\]

Note that the operator $P$ can be written in the form (cf. \eqref{3.5})
\begin{equation}
P=G(q,\varphi)\frac{d}{dx}+\frac{1}{2}\left[ G(q,\varphi)\right]_{x}
+Z(\psi)\frac{d^{3}}{dx^{3}},
\end{equation}
where
\begin{equation}
G^{ij}(q,\varphi) := b^{ij}_k q^k + (Z)^{ij}(\varphi)
=\sum_{m=0}^{n}\alpha_m G_{m}^{ij}(q,\varphi)
\end{equation}
is the most general flat Stäckel metric \cite{BS,BKM}.

\section{Multi-Hamiltonian hierarchies of evolutionary type}\label{S5}

The set of compatible Hamiltonian operators $P_{m}$ in \eqref{4.9} leads to
various multi-Hamiltonian hierarchies.

\subsection{Coupled Harry Dym hierarchy}\label{subs5.1}

First, we show that the set \eqref{4.9} contains known positive and negative
coupled Harry Dym (cHD) hierarchies \cite{AF1,AF2,BM1}. In order to fit the
notation to that known from the literature, let
\[
n=N,\qquad P_{m}=B_{N-m},\qquad q_{i}=u_{i},\qquad \psi^{1}=\psi,\qquad
\psi^{i}=\frac{1}{4}\varepsilon_{i},\quad i=1,\ldots,N-1\quad \text{and}\quad
\psi^{N}=0.
\]
Thus%
{\small\begin{equation}
B_{m}=\left(\!\renewcommand\arraystretch{1.2}
\begin{array}{c|c}%
\begin{matrix}%
 &  & J_{0}\\
 & \reflectbox{$\ddots$} & \vdots\\
J_{0} & \cdots & J_{m-1}%
\end{matrix}
& 0_{m\times(N-m)}\\\hline
0_{(N-m)\times m} &
\begin{matrix}%
-J_{m+1} & \cdots & -J_{N}\\
\vdots   & \reflectbox{$\ddots$} & \\
-J_{N}   &  & 
\end{matrix}
\end{array}
\!\right),\qquad m=0,\ldots,N,\label{5.1}%
\end{equation}}
where
\begin{align*}
J_{0}&=\varphi\partial+\psi\partial^{3},\qquad
J_{k}=\frac{1}{2}(u_{k}\partial+\partial u_{k})+\frac{1}{4}\varepsilon_{k}\partial^{3},
\quad k=1,\ldots,N-1,\\
J_{N}&=\frac{1}{2}(u_{N}\partial+\partial u_{N})=u_{N}^{\frac{1}{2}}\partial u_{N}^{\frac{1}{2}},
\qquad\text{and}\quad \partial \equiv\frac{\partial}{\partial x}.
\end{align*}
First, notice that the Casimir of $B_{0}$ is $C_{0}=\left(0,\ldots,0,au_{N}^{-\frac{1}{2}}\right)^{T}$
and the Casimir of $B_{N}$, which is $x$-independent, takes the form
$C_{N}=(c,0,\ldots,0)^{T}$, where $a$ and $c$ are arbitrary constants. Besides,
the operators $B_{m}$ satisfy the infinite recursion $B_{k+1}=\mathbf{R}B_{k}$,
$k\geqslant 0$, where all $B_{k}$ with $k>N$ are non-local and where the recursion
operator $\mathbf{R}$ and its inverse have the form%
\begin{equation}
\mathbf{R}=\left(\! \renewcommand\arraystretch{1.3}
\begin{array}{c|c}%
\begin{matrix}%
0 & \cdots & 0
\end{matrix}
& -J_{0}J_{N}^{-1}\\\hline%
\begin{matrix}%
1 &  & \\
 & \ddots & \\
 &  & 1
\end{matrix}
&
\begin{matrix}%
-J_{1}J_{N}^{-1}\\
\vdots\\
-J_{N-1}J_{N}^{-1}%
\end{matrix}
\end{array}
\!\right),\qquad
\mathbf{R}^{-1}=\left(\!\renewcommand\arraystretch{1.3}
\begin{array}{c|c}%
\begin{matrix}%
-J_{1}J_{0}^{-1}\\
\vdots\\
-J_{N-1}J_{0}^{-1}%
\end{matrix}
&
\begin{matrix}%
1 &  & \\
 & \ddots & \\
 &  & 1
\end{matrix}
\\\hline
J_{N}J_{0}^{-1} &
\begin{matrix}%
0 & \cdots & 0
\end{matrix}
\end{array}
\!\right).\label{5.2}%
\end{equation}
Then, the positive (local) cHD hierarchy has the form
\begin{equation}
\mathbf{u}_{t_{r}}=\mathbf{K}_{r}=\mathbf{R}^{r-1}\mathbf{K}_{1},
\qquad r=1,2,\ldots,\label{5.3}%
\end{equation}
where $\mathbf{u}=(u_{1},\ldots,u_{N})^{T}$ and for $a=1$
\begin{equation}
\mathbf{K}_{1}=B_{N}C_{0}=
\begin{pmatrix}
\left(u_{N}^{-\frac{1}{2}}\right)_{xxx}+\varphi\left(u_{N}^{-\frac{1}{2}}\right)_{x}\\
\frac{1}{4}\varepsilon_{1}\left(u_{N}^{-\frac{1}{2}}\right)_{xxx}
+u_{1}\left(u_{N}^{-\frac{1}{2}}\right)_{x}+\frac{1}{2}u_{N}^{-\frac{1}{2}}(u_{1})_{x}\\
\vdots\\
\frac{1}{4}\varepsilon_{N-1}\left(u_{N}^{-\frac{1}{2}}\right)_{xxx}
+u_{N-1}\left(u_{N}^{-\frac{1}{2}}\right)_{x}+\frac{1}{2}u_{N}^{-\frac{1}{2}}(u_{N-1})_{x}
\end{pmatrix}
.\label{5.4}%
\end{equation}
The negative (non-local) cHD hierarchy exists when $\varphi=0$ (so in
\eqref{4.8} there is no central extension of the first-order in this case) and
has the form
\begin{equation}
\mathbf{u}_{t_{-r}}=\mathbf{K}_{-r}=\mathbf{R}^{-r}\mathbf{K}_{0},
\qquad r=0,1,\ldots,\label{5.5}%
\end{equation}
where for $c=-2$
\begin{equation}
\mathbf{K}_{0}=B_{0}C_{N}=
\begin{pmatrix}
(u_{1})_{x}\\
(u_{2})_{x}\\
\vdots\\
(u_{N})_{x}
\end{pmatrix}
\label{5.6}%
\end{equation}
and where the first nontrivial vector field is
\begin{equation}
\mathbf{K}_{-1}=
\begin{pmatrix}
(u_{2})_{x}-\frac{1}{4}\varepsilon_{1}(u_{1})_{x}-u_{1}\partial^{-1}u_{1}-\frac{1}{2}(u_{1})_{x}\partial^{-2}u_{1}\\
\vdots\\
(u_{N})_{x}-\frac{1}{4}\varepsilon_{N-1}(u_{1})_{x}-u_{N-1}\partial^{-1}u_{1}-\frac{1}{2}(u_{N-1})_{x}\partial^{-2}u_{1}\\
-u_{N}\partial^{-1}u_{1}-\frac{1}{2}(u_{N})_{x}\partial^{-2}u_{1}
\end{pmatrix}
.\label{5.7}%
\end{equation}

\subsection{Coupled Korteweg-de Vries hierarchy}

Next, we show that the set \eqref{4.9} also contains known positive and
negative coupled Korteweg-de Vries (cKdV) hierarchies \cite{AF3,AF2,BM1}.
Again, in order to fit the notation to that known from the literature, let
\[
n=N,\qquad P_{m}=B_{m},\qquad q_{i}=-u_{N-i},\qquad
\psi^{i}=-\frac{1}{4}\varepsilon_{n-i},\quad i=1,\ldots,N,\qquad \psi^{0}=0\,.
\]
Then, Poisson tensors \eqref{4.9} take again the form \eqref{5.1} and the
recursion operator attains again the same form \eqref{5.2}, but now
\begin{equation}
J_{k}=\frac{1}{2}(u_{k}\partial+\partial u_{k})+\frac{1}{4}\varepsilon_{k}\partial^{3},
\qquad k=0,\ldots,N-1,\qquad J_{N}=-\varphi\partial.\label{5.8}%
\end{equation}
Notice that a Casimir of $B_{0}$ is now $C_{0}=(0,\ldots,0,c)^{T}$. The
positive (local) cKdV hierarchy has the form \eqref{5.3}, where
\[
\mathbf{u}=(u_{0},\ldots,u_{N-1})^{T},\qquad c=-2,\qquad \varphi=1,\qquad
\mathbf{K}_{1}=B_{N}C_{0}=
\begin{pmatrix}
(u_{0})_{x}\\
(u_{1})_{x}\\
\vdots\\
(u_{N-1})_{x}
\end{pmatrix}
\]
and the first nontrivial vector field is
\begin{align*}
\mathbf{K}_{2}&=\mathbf{R}\mathbf{K}_{1}
=\begin{pmatrix}
J_{0}u_{N-1}\\
(u_{0})_{x}+J_{1}u_{N-1}\\
\vdots\\
(u_{N-2})_{x}+J_{N-1}u_{N-1}
\end{pmatrix}\\
&=\begin{pmatrix}
\frac{1}{4}\varepsilon_{0}(u_{N-1})_{xxx}+u_{0}(u_{N-1})_{x}+\frac{1}{2}u_{N-1}(u_{0})_{x}\\
(u_{0})_{x}+\frac{1}{4}\varepsilon_{1}(u_{N-1})_{xxx}+u_{1}(u_{N-1})_{x}+\frac{1}{2}u_{N-1}(u_{1})_{x}\\
\vdots\\
(u_{N-2})_{x}+\frac{1}{4}\varepsilon_{N-1}(u_{N-1})_{xxx}+u_{N-1}(u_{N-1})_{x}+\frac{1}{2}u_{N-1}(u_{N-1})_{x}
\end{pmatrix}.
\end{align*}

The inverse (non-local) cKdV hierarchy exists when $\varepsilon_{0}=0$. Then,
the Casimir of $B_{N}$ is $C_{N}=(au_{0}^{-\frac{1}{2}},0,\ldots,0)^{T}$,
and for $a=-1$ the inverse hierarchy starts from
\[
\mathbf{K}_{-1}=B_{0}C_{N}=
\begin{pmatrix}
J_{1}u_{0}^{-\frac{1}{2}}\\
\vdots\\
J_{N-1}u_{0}^{-\frac{1}{2}}\\[4pt]
J_{N}u_{0}^{-\frac{1}{2}}
\end{pmatrix}
=
\begin{pmatrix}
\frac{1}{4}\varepsilon_{1}\left(u_{0}^{-\frac{1}{2}}\right)_{xxx}
+u_{1}\left(u_{0}^{-\frac{1}{2}}\right)_{x}+\frac{1}{2}u_{0}^{-\frac{1}{2}}(u_{1})_{x}\\
\vdots\\
\frac{1}{4}\varepsilon_{N-1}\left(u_{0}^{-\frac{1}{2}}\right)_{xxx}
+u_{N-1}\left(u_{0}^{-\frac{1}{2}}\right)_{x}+\frac{1}{2}u_{0}^{-\frac{1}{2}}(u_{N-1})_{x}\\[4pt]
-\left(u_{0}^{-\frac{1}{2}}\right)_{x}
\end{pmatrix}.
\]

\subsection{Triangular coupled Harry Dym hierarchy}

A third possibility occurs when we consider the operator $P_{n}$ in \eqref{4.9}
in the following way. In order to see the resemblance between the hierarchy
obtained below, which we will call the triangular cHD hierarchy, and the cHD
hierarchy above, let us use the following notation:
\[
n=N,\qquad q_{i}=-u_{N-i+1},\qquad i=1,\ldots,N.
\]
and choose $\psi^{i}=-\frac{1}{4}$, $i=0,\ldots,N$ (we recall that $Z_{n}=0$).
In the variables $\mathbf{u}=(u_{1},\ldots,u_{N})^{T}$ the operator $P_{n}$
becomes then
\begin{equation}
P_{N}=\frac{1}{4}
\begin{pmatrix}
 &  & 1\\
 & \reflectbox{$\ddots$} & \vdots\\
1 & \cdots & 1
\end{pmatrix}\partial^{3}
+\begin{pmatrix}
 &  & b_{1}\\
 & \reflectbox{$\ddots$} & \vdots\\
b_{1} & \cdots & b_{N}
\end{pmatrix},
\qquad b_{i}=\frac{1}{2}(u_{i}\partial+\partial u_{i})=u_{i}^{\frac{1}{2}}\partial u_{i}^{\frac{1}{2}}.
\label{5.10}%
\end{equation}
It is a sum of two compatible (due to the theory in the previous section)
Poisson operators
\begin{equation}
\pi_{0}=
\begin{pmatrix}
 &  & b_{1}\\
 & \reflectbox{$\ddots$} & \vdots\\
b_{1} & \cdots & b_{N}
\end{pmatrix},
\qquad
\pi_{1}=\frac{1}{4}
\begin{pmatrix}
 &  & 1\\
 & \reflectbox{$\ddots$} & \vdots\\
1 & \cdots & 1
\end{pmatrix}\partial^{3},
\label{5.11}%
\end{equation}
with the Casimir $C_{0}=(u_{1}^{-\frac{1}{2}},0,\ldots,0)^{T}$ and the
$x$-independent Casimir $C_{1}=(c_{1},\ldots,c_{N})^{T}$, respectively.

From these operators we can construct the following local hierarchy
\begin{equation}
\mathbf{u}_{t_{r}}=\mathbf{K}_{r}=\mathbf{R}^{r-1}\mathbf{K}_{1},
\qquad r=1,2,\ldots
\label{5.12}%
\end{equation}
where
\begin{equation}
\mathbf{R}=\pi_{1}\pi_{0}^{-1},\qquad
\mathbf{K}_{1}=\pi_{1}C_{0}=
\begin{pmatrix}
0\\ \vdots\\ 0\\ \frac{1}{4}(u_{1}^{-\frac{1}{2}})_{xxx}
\end{pmatrix},
\label{5.13}%
\end{equation}
and the following non-local hierarchy
\begin{equation}
\mathbf{u}_{t_{-r}}=\mathbf{K}_{-r}=\mathbf{R}^{1-r}\mathbf{K}_{-1},
\qquad r=1,2,\ldots,
\label{5.14}%
\end{equation}
where
\begin{equation}
\mathbf{R}^{-1}=\pi_{0}\pi_{1}^{-1},\qquad
\mathbf{K}_{-1}=\pi_{0}C_{1}=
\begin{pmatrix}
(u_{1})_{x}\\
(u_{1})_{x}+(u_{2})_{x}\\
\vdots\\
\sum_{i=1}^{N}(u_{i})_{x}
\end{pmatrix}.
\label{5.15}%
\end{equation}
For $N=1$ it yields standard local and non-local Harry Dym (HD) hierarchies
with $u_{1}=u$, $\pi_{0}=u^{\frac{1}{2}}\partial u^{\frac{1}{2}}$ and
$\pi_{1}=\frac{1}{4}\partial^{3}$, with the first flows given by
\begin{align}
K_{1} &=\pi_{1}C_{0}=\frac{1}{4}(u^{-\frac{1}{2}})_{xxx},\nonumber\\
K_{2} &=RK_{1}=-\frac{1}{16}\bigl(u^{-\frac{1}{2}}(u^{-\frac{1}{2}})_{x}^{2}\bigr)_{xxx}
=-\frac{1}{64}(u^{-\frac{7}{2}}u_{x}^{2})_{xxx},\nonumber\\
&\vdots\nonumber
\end{align}
and by ($c_{1}=2$)
\begin{align}
K_{-1} &=u_{x},\nonumber\\
K_{-2} &=R^{-1}K_{-1}=2u_{x}\partial^{-2}u+4u\partial^{-1}u,\nonumber\\
&\vdots\nonumber
\end{align}
respectively. For $N>1$ the local hierarchy is degenerate, as the matrix of
$\mathbf{R}$ has zeros over the diagonal and the recursion operator of HD on
the diagonal so that the local hierarchy of vector fields takes the form
$\mathbf{K}_{r}=(0,\ldots,0,K_{r}[u_{1}])^{T}$. On the other hand, the non-local
hierarchy has a triangular non-local coupled Harry Dym form, as
\begin{equation}
(\mathbf{R}_{ij})^{-1}=
\begin{cases}
R_{1}^{-1},& i=j,\\
R_{i-j+1}^{-1}-R_{i-j}^{-1},& i>j,\\
0,& i<j,
\end{cases}
\label{5.17}%
\end{equation}
where $R_{k}^{-1}=4u_{k}^{\frac{1}{2}}\partial u_{k}^{\frac{1}{2}}\partial^{-3}
=4u_{k}\partial^{-2}+2(u_{k})_{x}\partial^{-3}$. Then, the components of the
first two flows are given by
\begin{equation}
(\mathbf{K}_{-1})_{i}=\sum_{j=1}^{i}(u_{j})_{x},\qquad
(\mathbf{K}_{-2})_{i}=\sum_{j=1}^{i}R_{j}^{-1}(u_{i-j+1})_{x},\qquad
 i=1,\ldots,N,
\label{5.18}%
\end{equation}
so that
\[
\mathbf{K}_{-1}
=\begin{pmatrix}
(u_{1})_{x}\\
(u_{1}+u_{2})_{x}\\
(u_{1}+u_{2}+u_{3})_{x}\\
\vdots\\
(u_{1}+u_{2}+\ldots+u_{N})_{x}
\end{pmatrix},
\qquad \text{and}
\]
\[
\mathbf{K}_{-2}
=\begin{pmatrix}
2(u_{1})_{x}\partial^{-2}u_{1}+4u_{1}\partial^{-1}u_{1}\\
2\bigl[(u_{1})_{x}\partial^{-2}u_{2}+(u_{2})_{x}\partial^{-2}u_{1}\bigr]
+4\bigl[u_{1}\partial^{-1}u_{2}+u_{2}\partial^{-1}u_{1}\bigr]\\
2\bigl[(u_{1})_{x}\partial^{-2}u_{3}+(u_{2})_{x}\partial^{-2}u_{2}+(u_{3})_{x}\partial^{-2}u_{1}\bigr]
+4\bigl[u_{1}\partial^{-1}u_{3}+u_{2}\partial^{-1}u_{2}+u_{3}\partial^{-1}u_{1}\bigr]\\
\vdots\\
(\mathbf{K}_{-2})_{N}
\end{pmatrix}.
\]

The non-local hierarchy \eqref{5.14} with the bi-Hamiltonian structure given by the
Poisson operators \eqref{5.11} is a special case of the multi-component hierarchy
of the Camassa-Holm type constructed in \cite{SS} with respect to the Novikov
algebras $\mathbb{T}_n$, which coincide with the $n$-dimensional algebras
$\mathcal{A}^n$ defined by \eqref{algm} for $m=n$. Also, this hierarchy can be obtained from the formula for the densities in Example I in \cite{Kon}, in the context of the same algebra $\mathcal{A}^n$.

\subsection{Triangular coupled Korteweg-de Vries hierarchy}

The last possibility occurs if we take again the Poisson tensor $P_{N}$ in \eqref{5.10}
\begin{equation}\label{p1}
P_{N}=\pi_{1}=\frac{1}{4}
\begin{pmatrix}
 &  & 1\\
 & \reflectbox{$\ddots$} & \vdots\\
1 & \cdots & 1
\end{pmatrix}\partial^{3}
+\begin{pmatrix}
 &  & b_{1}\\
 & \reflectbox{$\ddots$} & \vdots\\
b_{1} & \cdots & b_{N}
\end{pmatrix}
\end{equation}
and the first order Poisson tensor
\begin{equation}\label{p0}
\pi_{0}=
\begin{pmatrix}
 &  & 1\\
 & \reflectbox{$\ddots$} & \vdots\\
1 & \cdots & 1
\end{pmatrix}\partial,
\end{equation}
compatible with $\pi_{1}$ since $\pi_{0}$ is the first-order central extension
of the operator $\Pi_{n}$. Only $\pi_{0}$ has a local Casimir of the form
$(c_{1},\ldots,c_{N})^{T}$. It generates a local hierarchy, the triangular
coupled KdV hierarchy of the form
\begin{equation}
\mathbf{u}_{t_{r}}=\mathbf{K}_{r}=\mathbf{R}^{r-1}\mathbf{K}_{1},
\qquad r=1,2,\ldots,\label{6.26}%
\end{equation}
where
\begin{equation}
\mathbf{u}=(u_{1},\ldots,u_{N})^{T},\qquad
\mathbf{R}=\pi_{1}\pi_{0}^{-1},\qquad
\mathbf{K}_{1}=\pi_{1}C_{0}=
\begin{pmatrix}
(u_{1})_{x}\\
(u_{1})_{x}+(u_{2})_{x}\\
\vdots\\
\sum_{i=1}^{N}(u_{i})_{x}
\end{pmatrix},
\label{6.27}%
\end{equation}
with the choice $c_{i}=2$, $i=1,\ldots,N$ and
\begin{equation}
\mathbf{R}_{ij}=
\begin{cases}
R_{1},&i=j,\\
R_{i-j+1}-R_{i-j},& i>j,\\
0,& i<j,
\end{cases}
\qquad
R_{k}=\frac{1}{4}\partial^{2}+u_{k}+\frac{1}{2}(u_{k})_{x}\partial^{-1}.
\label{6.28}%
\end{equation}
Then, the components of the first two flows of this hierarchy are
\begin{equation}
(\mathbf{K}_{1})_{i}=\sum_{j=1}^{i}(u_{j})_{x},\qquad
(\mathbf{K}_{2})_{i}=\sum_{j=1}^{i}R_{j}(u_{i-j+1})_{x},\qquad
\ldots,\qquad i=1,\ldots,N,
\label{6.29}%
\end{equation}
so that
\[
\mathbf{K}_{1}=
\begin{pmatrix}
(u_{1})_{x}\\
(u_{1}+u_{2})_{x}\\
(u_{1}+u_{2}+u_{3})_{x}\\
\vdots\\
(u_{1}+u_{2}+\ldots+u_{N})_{x}
\end{pmatrix},
\quad
\mathbf{K}_{2}=
\begin{pmatrix}
\frac{1}{4}(u_{1})_{xxx}+\frac{3}{2}u_{1}(u_{1})_{x}\\
\frac{1}{4}(u_{1}+u_{2})_{xxx}+\frac{3}{2}(u_{1}u_{2})_{x}\\
\frac{1}{4}(u_{1}+u_{2}+u_{3})_{xxx}+\frac{3}{2}\bigl(u_{1}u_{3}+\frac{1}{2}u_{2}^{2}\bigr)_{x}\\
\vdots\\
(\mathbf{K}_{2})_{N}
\end{pmatrix},
\quad \ldots
\]

The local hierarchy \eqref{6.26}, with the bi-Hamiltonian
structure given by the Poisson operators \eqref{p1} and \eqref{p0}, is
another special case of the same multi-component Camassa-Holm type hierarchy
from \cite{SS}, mentioned at the end of the previous subsection.

The triangular hierarchies constructed in Subsections~3 and~4 are generated by
the Novikov algebra $\mathcal{A}^{n}$, where the multiplication matrix $A_{n}$
contains all basic elements $e^{i}$. For the remaining Novikov algebras
$\mathcal{A}^{m}$, $m=0,\ldots,n-1$, the multiplication matrices $A_{m}$ do not
contain all basic elements $e^{i}$ and, consequently, the constructed
triangular systems are degenerate and thus not interesting.

\setcounter{equation}{0}
\renewcommand{\theequation}{A.\arabic{equation}}

\section{Appendix A}

\subsection*{Proof of Theorem \ref{Z}}

Assume first that all the indices $i,j,k\in I_{1}^{m}$. Then, given
\eqref{3.1}, the left-hand side of \eqref{3.3} reads%
\begin{align*}
\sum\limits_{s=0}^{n-m-1}\Bigl( \delta_{s}^{i+j+m-n-1}(Z_{m})^{sk}-\delta
_{s}^{j+k+m-n-1}(Z_{m})^{is}\Bigr)
&=(Z_{m})^{i+j+m-n-1,k}-(Z_{m})^{i,j+k+m-n-1}\\
\overset{\eqref{3.4}}{=}
\varphi_{m}^{i+j+k+2m-2n-2} - \varphi_{m}^{i+j+k+2m-2n-2}
&=0,
\end{align*}
due to the fact that in this case $i+j+m-n-1\leqslant n-m-1$. Similar
calculations show that \eqref{3.3} is satisfied in the case $i,j,k\in
I_{2}^{m}$:
\begin{align*}
\sum\limits_{s=n-m+1}^{n}\Bigl( \delta_{s}^{i+j+m-n-1}(Z_{m})^{sk}-\delta
_{s}^{j+k+m-n-1}(Z_{m})^{is}\Bigr)
&=(Z_{m})^{i+j+m-n-1,k}-(Z_{m})^{i,j+k+m-n-1}\\
\overset{\eqref{3.4}}{=}
\varphi_{m}^{i+j+k+2m-2n-2}-\varphi_{m}^{i+j+k+2m-2n-2}
&=0,
\end{align*}
due to the fact that in this case $n-m+1\leqslant i+j+m-n-1$. Finally, let us
assume that one of the indices, say $k$, belongs to the other index set.
Assume thus that $i,j\in I_{1}^{m}$ while $k\in I_{2}^{m}$ (all other such
situations are proved analogously). Then, given \eqref{3.1}, the first sum on
the left-hand side of \eqref{3.3} is%
\[
(b_{m})_{s}^{ij}(Z_{m})^{sk}=\sum\limits_{s=n-m+1}^{n}%
\delta_{s}^{i+j+m-n-1}(Z_{m})^{sk}=0
\]
as the index $i+j+m-n-1\leqslant n-m-1$ for $i,j\in I_{1}^{m}$. The second sum on
the left-hand side of \eqref{3.3} is
\[
\sum\limits_{s=0}^{n-m-1}(b_{m})_{s}^{jk}(Z_{m})^{is}
\]
and it is also $0$ since $j\in I_{1}^{m}$ and $k\in I_{2}^{m}$. Finally, a
direct computation shows that under the symmetry assumption $Z^{pq}=Z^{qp}$
the matrix of the system \eqref{3.3} has co-rank $n$ and thus \eqref{3.4}
is the ($n$-parameter) general solution of \eqref{3.3}.

\subsection*{Proof of Lemma \ref{L7}}

\bigskip
The condition \eqref{war} reads (again, no summation over $m$)
\begin{equation}
(b_{m})_{k}^{ij}(Z_{m})^{ks}=0\quad \text{for}\quad i,j,s=1,\ldots,n.
\label{wk}%
\end{equation}
We will now show that it necessarily implies that $Z_{m}=0$. For a fixed $m$
and any $s$, suppose that $i,j\in I_{1}^{m}$. Then, \eqref{wk} reads
\[
0=(b_{m})_{k}^{ij}(Z_{m})^{ks}=\sum_{k=1}^{n}\delta_{k}^{i+j+m-n-1}%
(Z_{m})^{ks}=(Z_{m})^{i+j+m-n-1,s},
\]
which implies $(Z_{m})^{\alpha s}=0$ for $1\leqslant \alpha\leqslant n-m-1$ (and all $s$)
since in this case $1+m-n\leqslant i+j+m-n-1\leqslant n-m-1$. Further, for $i,j\in
I_{2}^{m}$, \eqref{wk} reads%
\[
0=(b_{m})_{k}^{ij}(Z_{m})^{ks}=-\sum_{k=1}^{n}\delta_{k}^{i+j+m-n-1}%
(Z_{m})^{ks}=-(Z_{m})^{i+j+m-n-1,s},
\]
which implies that $(Z_{m})^{\alpha s}=0$ for $n-m+1\leqslant \alpha\leqslant n+m-1$
(and all $s$), since in this case $n-m+1\leqslant i+j+m-n-1\leqslant n+m-1$. Thus, in
$(Z_{m})^{\alpha s}$ all the rows except the row $n-m$ (and, in the case
$m=0$, the row $n$) vanish. Since $Z_{m}$ is antisymmetric, it follows that
$Z_{m}=0$. So, no nontrivial $Z_{m}$ exist that yield an order $2$ cocycle.

\subsection*{Proof of Lemma \ref{L2}}

The associativity condition \eqref{compat} is equivalent to
\begin{equation}\label{asoc}
(e^{i}\circ_m e^{j})\circ_p e^{k} + (e^{i}\circ_p e^{j})\circ_m e^{k}
= e^{i}\circ_m (e^{j}\circ_p e^{k}) + e^{i}\circ_p (e^{j}\circ_m e^{k}),
\end{equation}
for all $i,j,k=1,\ldots,n$ and for all $m,p=0,\ldots,n$. Explicitly, the
condition \eqref{asoc} reads
\begin{equation}\label{asoc2}
(b_m)^{ij}_s(b_p)^{sk}_r+(b_p)^{ij}_s(b_m)^{sk}_r
-(b_m)^{is}_r(b_p)^{jk}_s-(b_p)^{is}_r(b_m)^{jk}_s=0,
\end{equation}
for all $i,j,k,r=1,\ldots,n$ and for all $m,p=0,\ldots,n$.
For $m=p$ this lemma reduces to Lemma \ref{L1}. Assume thus that $m>p$ (so that
$n-m<n-p$). There are, up to permutations and analogous situations, only four
different cases:
1. $i,j,k\in I_{1}^{m}=\{1,\ldots,n-m\}\subset I_{1}^{p}$,
2. $i,j\in I_{1}^{m}=\{1,\ldots,n-m\}$, $k\in I_{1}^{p}\cap I_{2}^{m}=\{n-m+1,\ldots,n-p\}$,
3. $i\in I_{1}^{m}$, $j\in I_{1}^{p}\cap I_{2}^{m}$, $k\in I_{2}^{p}\subset I_{2}^{m}$,
and 4. $i\in I_{1}^{m}$, $j,k\in I_{1}^{p}\cap I_{2}^{m}$. In the case 1 all
terms in \eqref{asoc2} are equal so it is satisfied. For example, the first
term becomes
\begin{align}
(b_{m})_{s}^{ij}(b_{p})_{r}^{sk}
={\textstyle\sum\limits_{s=1}^{n}}\delta_{s}^{i+j+m-n-1}\delta_{r}^{s+k+p-n-1}
=\delta_{r}^{i+j+k+m+p-2n-2}.
\end{align}
All the other terms in \eqref{asoc2} yield the same expression as all the
indices $i,j,k\in I_{1}^{m}\cap I_{1}^{p}$. In the case 2, the first term in
\eqref{asoc2} reads
\[
(b_{m})_{s}^{ij}(b_{p})_{r}^{sk}=
{\textstyle\sum\limits_{s=1}^{n-p}}
\delta_{s}^{i+j+m-n-1}\delta_{r}^{s+k+p-n-1}
=\delta_{r}^{i+j+k+m+p-2n-2}
\]
and is equal exactly to the third term, as the third term becomes
\[
(b_m)^{is}_r(b_p)^{jk}_s=
{\textstyle\sum\limits_{s=1}^{n-m}}
\delta_{r}^{i+s+m-n-1}\delta_{s}^{j+k+p-n-1}
=\delta_{r}^{i+j+k+m+p-2n-2}.
\]
The second term $(b_{p})_{s}^{ij}(b_{m})_{r}^{sk}$ is equal to zero, as
$(b_{p})_{s}^{ij}=\delta_{s}^{i+j+p-n-1}$ is non-zero only for
$s=i+j+p-n-1\in I_{1}^{m}$ and then $(b_{m})_{r}^{sk}$ becomes zero as
$k\in I_{2}^{m}$. For analogous reasons, the fourth term is also zero. Thus,
also in this case \eqref{asoc2} is satisfied. In the case 3 all four terms are
zero. The reason is as follows.

The first term $(b_m)^{ij}_s(b_p)^{sk}_r$ is zero since $(b_m)^{ij}_s=0$ as
$i\in I_{1}^{m}$ while $j\in I_{2}^{m}$.

The second term $(b_p)^{ij}_s(b_m)^{sk}_r$ is zero as $(b_p)^{ij}_s=
\delta_{s}^{i+j+p-n-1}$ is nonzero only if $s=i+j+p-n-1<n-m$, i.e. only if
$s\in I_{1}^{m}$, and then $(b_m)^{sk}_r=0$ since $k\in I_{2}^{p}\subset
I_{2}^{m}$.

The third term $(b_m)^{is}_r(b_p)^{jk}_s$ is zero since $(b_p)^{jk}_s=0$ as
$j\in I_{1}^{p}$ while $k\in I_{2}^{p}$.

The fourth term $(b_p)^{is}_r(b_m)^{jk}_s$ is zero as
\[
(b_p)^{is}_r(b_m)^{jk}_s
=-{\textstyle\sum\limits_{s=1}^{n-m}}\delta_{r}^{i+s+p-n-1}\delta_{s}^{j+k+m-n-1}=0,
\]
since $j+k+m-n-1\geqslant n-p+1>n-m+1$ so the sum disappears. So, also in the case 3
the formula \eqref{asoc2} is satisfied. Finally, let us analyze the case 4.
The first term $(b_m)^{ij}_s(b_p)^{sk}_r$ is zero since $i\in I_{1}^{m}$ while
$j\in I_{2}^{m}$ so $(b_m)^{ij}_s=0$. The second term $(b_p)^{ij}_s(b_m)^{sk}_r$
is zero, as
\[
(b_p)^{ij}_s(b_m)^{sk}_r
=-{\textstyle\sum\limits_{s=n-m+1}^{n}}\delta_{s}^{i+j+p-n-1}\delta_{r}^{s+k+m-n-1}=0
\]
due to the fact that $i+j+p-n-1<n-m+1$ so that the sum disappears. Finally,
the last two terms in \eqref{asoc2} cancel each other. The third term is
\[
(b_m)^{is}_r(b_p)^{jk}_s
={\textstyle\sum\limits_{s=1}^{n-m}}\delta_{r}^{i+s+m-n-1}\delta_{s}^{j+k+p-n-1}
=\delta_{r}^{i+j+k+m+p-2n-2},
\]
while the fourth term is
\[
(b_p)^{is}_r(b_m)^{jk}_s
=-{\textstyle\sum\limits_{s=1}^{n-p}}\delta_{r}^{i+s+p-n-1}\delta_{s}^{j+k+m-n-1}
=-\delta_{r}^{i+j+k+m+p-2n-2},
\]
so they cancel each other. Thus, the lemma is proved.

\subsection*{Proof of Theorem \ref{Z1}}

Given \eqref{4.4}, \eqref{2.9} can be rewritten as%
\begin{align*}
0&=Z\bigl(\alpha_{p}(b_{p})_{s}^{ij}e^{s},e^{k}\bigr)
 -Z\bigl(e^{i},\alpha_{p}(b_{p})_{s}^{jk}e^{s}\bigr)
 =\sum\limits_{p=0}^{n}\alpha_{p}\Bigl[(b_{p})_{s}^{ij}Z(e^{s},e^{k})
 -(b_{p})_{s}^{jk}Z(e^{i},e^{s})\Bigr]\\
&=\sum\limits_{m=0}^{n}\sum\limits_{p=0}^{n}\alpha_{p}\alpha_{m}\Bigl[
(b_{p})_{s}^{ij}(Z_{m})^{sk}-(b_{p})_{s}^{jk}(Z_{m})^{is}\Bigr]\\
&=\sum\limits_{m=0}^{n}\alpha_{m}^{2}\,
\overset{\text{cancels by Theorem \ref{Z}}}{\overbrace{\Bigl[
(b_{m})_{s}^{ij}(Z_{m})^{sk}-(b_{m})_{s}^{jk}(Z_{m})^{is}\Bigr]}}\\
&\quad+\sum\limits_{m=0}^{n}\sum\limits_{m>p}\alpha_{m}\alpha_{p}\Bigl[
(b_{p})_{s}^{ij}(Z_{m})^{sk}-(b_{p})_{s}^{jk}(Z_{m})^{is}
+(b_{m})_{s}^{ij}(Z_{p})^{sk}-(b_{m})_{s}^{jk}(Z_{p})^{is}\Bigr].
\end{align*}

Due to \eqref{3.1} it is equivalent to the condition
\begin{equation}
(b_{p})_{s}^{ij}(Z_{m})^{sk}-(b_{p})_{s}^{jk}(Z_{m})^{is}
+(b_{m})_{s}^{ij}(Z_{p})^{sk}-(b_{m})_{s}^{jk}(Z_{p})^{is}=0.
\label{4.7}%
\end{equation}
for all $m,p=0,\ldots,n$ and all $i,j,k=1,\ldots,n$ (the index $s$ varies from
$1$ to $n$). Note that the condition in Remark \ref{Bl} yields directly the
formula \eqref{4.7}. For $m=p$ this formula reduces to \eqref{3.3} in the
context of Theorem \ref{Z} and has been proved above in this appendix. Let us
thus assume that $m>p$. There are, up to permutations and analogous
situations, only three different cases: 1. $i,j,k\in I_{1}^{m}=\{1,\ldots,n-m\}$,
2. $i,j\in I_{1}^{m}=\{1,\ldots,n-m\}$, $k\in I_{1}^{p}\cap I_{2}^{m}=\{n-m+1,\ldots,n-p\}$,
and 3. $i\in I_{1}^{m}$, $j\in I_{1}^{p}\cap I_{2}^{m}$, $k\in I_{2}^{p}$.

Consider the case 1: $i,j,k\in I_{1}^{m}=\{1,\ldots,n-m\}$. Let us calculate
separately all the terms in \eqref{4.7}:%
\begin{equation}
(b_{p})_{s}^{ij}(Z_{m})^{sk}
=\sum\limits_{s=1}^{n-m}\delta_{s}^{i+j+p-n-1}(Z_{m})^{sk}=(Z_{m})^{i+j+p-n-1,k}
=\varphi^{i+j+k+m+p-2n-2},
\end{equation}
since $k\in I_{1}^{m}$ and $(i+j+p-n-1)\in I_{1}^{m}$ as
$i+j+p-n-1\leqslant n-2m+p-1<n-m-1$,%
\begin{align}
(b_{p})_{s}^{jk}(Z_{m})^{is}
=\sum\limits_{s=1}^{n-m}\delta_{s}^{j+k+p-n-1}(Z_{m})^{is}=(Z_{m})^{i,j+k+p-n-1}
=\varphi^{i+j+k+p+m-2n-2},
\end{align}
since $i\in I_{1}^{m}$ and $(j+k+p-n-1)\in I_{1}^{m}$ as
$j+k+p-n-1\leqslant n-2m+p-1<n-m-1$,%
\begin{align}
(b_{m})_{s}^{ij}(Z_{p})^{sk}
=\sum\limits_{s=1}^{n-m}\delta_{s}^{i+j+m-n-1}(Z_{p})^{sk}=(Z_{p})^{i+j+m-n-1,k}
=\varphi^{i+j+k+m+p-2n-2},
\end{align}
since $k\in I_{1}^{m}\subset I_{1}^{p}$ and $(i+j+m-n-1)\in I_{1}^{p}$ as
$i+j+m-n-1\leqslant n-m-1\leqslant n-p$, and finally
\begin{align}
(b_{m})_{s}^{jk}(Z_{p})^{is}
=\sum\limits_{s=1}^{n-m}\delta_{s}^{j+k+m-n-1}(Z_{p})^{is}=(Z_{p})^{i,j+k+m-n-1}
=\varphi^{i+j+k+m+p-2n-2},
\end{align}
since $i\in I_{1}^{m}\subset I_{1}^{p}$ and $j+k+m-n-1\in I_{1}^{p}$ as
$j+k+m-n-1\leqslant n-m-1\leqslant n-p$. In consequence, all terms in \eqref{4.7} are
equal (the same is true when $i,j,k\in I_{2}^{p}=\{n-p+1,\ldots,n\}$) and thus
\eqref{4.7} is satisfied.

Consider now the case 2: $i,j\in I_{1}^{m}=\{1,\ldots,n-m\}$, $k\in
I_{1}^{p}\cap I_{2}^{m}=\{n-m+1,\ldots,n-p\}$. Again, let us calculate each
term in \eqref{4.7}:%
\[
(b_{p})_{s}^{ij}(Z_{m})^{sk}\overset{I_{1}^{m}\subset I_{1}^{p}}{=}
\sum\limits_{s=n-m+1}^{n}\delta_{s}^{i+j+p-n-1}(Z_{m})^{sk}=0
\]
since $k\in I_{2}^{m}$ and $i+j+p-n-1\in I_{1}^{m}$ as
$i+j+p-n-1\leqslant n-2m+p-1<n-m-1,$%
\begin{align}
(b_{p})_{s}^{jk}(Z_{m})^{is}
\overset{I_{1}^{m}\subset I_{1}^{p}}{=}\sum\limits_{s=1}^{n-m}\delta_{s}^{j+k+p-n-1}(Z_{m})^{is}
=(Z_{m})^{i,j+k+p-n-1}
=\varphi^{i+j+k+p+m-2n-2},
\end{align}
since $i\in I_{1}^{m}$ and $(j+k+p-n-1)\in I_{1}^{m}$ as $j+k+p-n-1\leqslant n-m-1,$%
\begin{align}
(b_{m})_{s}^{ij}(Z_{p})^{sk}
\overset{s,k\in I_{1}^{p}}{=}\sum\limits_{s=1}^{n-p}\delta_{s}^{i+j+m-n-1}(Z_{p})^{sk}
=(Z_{p})^{i+j+m-n-1,k}
=\varphi^{i+j+k+p+m-2n-2},
\end{align}
since $k\in I_{1}^{p}$ and $(i+j+m-n-1)\in I_{1}^{p}$ as $i+j+m-n-1\leqslant n-m-1$,
and finally
\[
(b_{m})_{s}^{jk}(Z_{p})^{is}=0\quad \text{since}\quad j\in I_{1}^{m}\ \text{and}\ k\in I_{2}^{m}.
\]
Thus, the first and the fourth terms in \eqref{4.7} are equal to zero while
the middle terms cancel each other. Hence, \eqref{4.7} is satisfied also in
case~2.

Finally, consider the case 3: $i\in I_{1}^{m}=\{1,\ldots,n-m\}$,
$j\in I_{1}^{p}\cap I_{2}^{m}=\{n-m+1,\ldots,n-p\}$, $k\in I_{2}^{p}=\{n-p+1,\ldots,n\}$.
Then,
\[
(b_{p})_{s}^{ij}(Z_{m})^{sk}\overset{k\in I_{2}^{p}\subset I_{2}^{m}}{=}
\sum\limits_{s=n-m+1}^{n}\delta_{s}^{i+j+p-n-1}(Z_{m})^{sk}=0
\]
since $k\in I_{2}^{p}$ and $(i+j+p-n-1)\in I_{1}^{m}\subset I_{1}^{p}$ as
$i+j+p-n-1\leqslant n-m-1<n-p,$
\[
(b_{p})_{s}^{jk}(Z_{m})^{is}=0\quad \text{since}\quad j\in I_{1}^{p}\ \text{and}\ k\in I_{2}^{p},
\]
\[
(b_{m})_{s}^{ij}(Z_{p})^{sk}=0\quad \text{since}\quad i\in I_{1}^{m}\ \text{and}\ j\in I_{2}^{m},
\]
and, finally
\[
(b_{m})_{s}^{jk}(Z_{p})^{is}\overset{i\in I_{1}^{m}\subset I_{1}^{p}}{=}
-\sum\limits_{s=1}^{n-m}\delta_{s}^{j+k+m-n-1}(Z_{p})^{is}=0,
\]
since $i\in I_{1}^{m}$ and $(j+k+m-n-1)\in I_{2}^{m}$ as
$j+k+m-n-1\geqslant n-p+1>n-m+1$. Thus, in this case all the terms in \eqref{4.7}
vanish and therefore \eqref{4.7} is satisfied. This means that the pencil
\eqref{4.4} with $Z_{m}$ given by \eqref{4.6} is a particular solution of the
Frobenius condition \eqref{3.3}. The theorem is proved.

\begin{small}

\end{small}

\end{document}